\documentclass[11pt]{article}

\usepackage{amsmath}
\usepackage[unicode]{hyperref}
\usepackage{caption}
\usepackage{color}

\usepackage{graphicx,isolatin1}
\usepackage[squaren]{SIunits}
\usepackage{etoolbox}

\newcommand{\gfrac}[2]{\displaystyle\frac{#1}{#2}}
\newcommand{\dd}{\mbox{d}}

\DeclareMathOperator\arctanh{arctanh}

\begin{document}

\title{Pair invariant mass spectrum and polarization asymmetry in the event generation of gamma-ray conversions
}

\author{D.~Bernard
\\
LLR, Ecole Polytechnique, CNRS/IN2P3, 91128 Palaiseau, France}

\maketitle

\begin{abstract}
We examine the dielectron invariant mass spectrum and the variation of the
polarization asymmetry as a function of that mass in the event
generator of the five-dimensional Bethe-Heitler differential cross
section of the conversion of linearly polarized photons
that we have implemented recently as a Geant4 Physics Model.
We compare the results obtained with simulated samples to analytical
expressions.

Studies of $b \to s \gamma$ decays at LHC and Belle II experiments
could provide hints of physics beyond the standard model in
flavor-changing neutral current processes; issues have been reported in
the simulation of the properties of the background induced
by genuine $\gamma$-ray conversions in their detectors.
Our results demonstrate that using the five-dimensional model
would help solve these issues.
\end{abstract}
 
{\em keywords }:
gamma rays,  
pair conversion,  
polarization, 
radiative decays, 
dielectron, 
Geant4

\section{ Motivation}
\label{sec:Motivation}

The standard model of elementary particles and of their interactions
(SM) predicts that $b \to s \gamma$ decays produce photons that are
left-handed to a high accuracy
\cite{Atwood:1997zr}.
Photon polarimetry offers therefore a unique sensitivity to 
beyond-the-standard-model (BSM) processes that provide right-handed couplings.
 
Among the various methods that have been used or that are being
developed to achieve the measurement, the analysis can involve the
angular distribution of the final state including a prompt conversion
of the photon to an $\ell^+\ell^-$ pair at the $B$ decay vertex
(``short distance lepton pairs''),
or the
``nuclear'' conversion of a real photon in the material of the
detector far away from the vertex 
(``long distance lepton pairs''),
\cite{Bishara:2015yta,Grossman:2000rk}.
In the later case, the system formed by the two vector particles
($K^*$ and $\gamma$) stay in a coherent state until decay and
conversion.

In both cases (short distance, discussion of eq. (2.15) of
\cite{Grossman:2000rk}; long distance, \cite{Borsellino:1953zz}) most
of the sensitivity lies at low values of the lepton pair invariant
mass, $\mu$, therefore the analysis is performed as a function of
$\mu$ and light-lepton (electron-positron) final states are used.
In that context, understanding the $\mu$ spectrum and
polarization asymmetry of the conversion of background photons in the
detector can be critical to mastering the systematics;
the examination of event samples obtained by simulation with the
Geant4 toolkit \cite{Allison:2016lfl} showed
\cite{Aaij:2015dea,Borsato:2015zuf} a departure from the QED
prediction \cite{Borsellino:1953zz}.
The physics models that were existing in Geant4 at that time were
sampling probability density functions (pdf) that are products of
one-dimensional Erzätze, with a focus on the distributions of the
one-track polar angle and on the energy share between the electron and
the positron, but with an inconsistent behavior for other variables 
\cite{Gros:2016zst}.
In particular the polarized model was found to be flawed
\cite{Gros:2016zst}.

In this paper we examine the dielectron invariant mass spectrum and the
variation of the polarization asymmetry as a function of that mass in
the event generator of the five-dimensional Bethe-Heitler differential
cross section of the conversion of linearly polarized photons.
We compare the results obtained with simulated samples to analytical
expressions.

\section{Dielectron invariant mass spectrum}

We examine the $\mu$ spectra provided by the exact, that is,
five-dimensional, event generator of the Bethe-Heitler conversion
process that we have recently implemented in Geant4
\cite{Bernard:2018hwf,Semeniouk:2019cwl,Bernard:2019qvd,Ivanchenko:2019ovi},
and we compare it with the Bethe-Heitler analytical spectrum obtained
by Borsellino \cite{Borsellino:1953zz}.
Note that in contrast with Bethe-Heitler \cite{Bethe-Heitler}, the
algorithm used here does not assume that the energy
carried away by the recoiling target be negligible (compare the
differential element of eq.~(20) of \cite{Bethe-Heitler} to that of
eqs. (1)-(3) of \cite{Bernard:2018hwf}).

The $\mu$ spectrum shows a maximum at very low mass
($\mu < 2\,\mega\electronvolt/c^2$) and then decreases as $1/\mu^3$
\cite{Borsellino:1953zz}, so using the low-mass approximation
expression from \cite{Borsellino:1953zz} is legitimate
($\mu \ll E$, where $E$ is the energy of the incident photon).
Indeed for the nuclear conversion of 40\,GeV photons on silicon, 
the fraction of the spectrum above $\mu = 1\,\giga\electronvolt/c^2$ 
is found to be as small as $2 \times 10^{-5}$ in the simulated samples.

\begin{figure}[ht!]
\begin{center}
\includegraphics[width=0.47\textwidth]{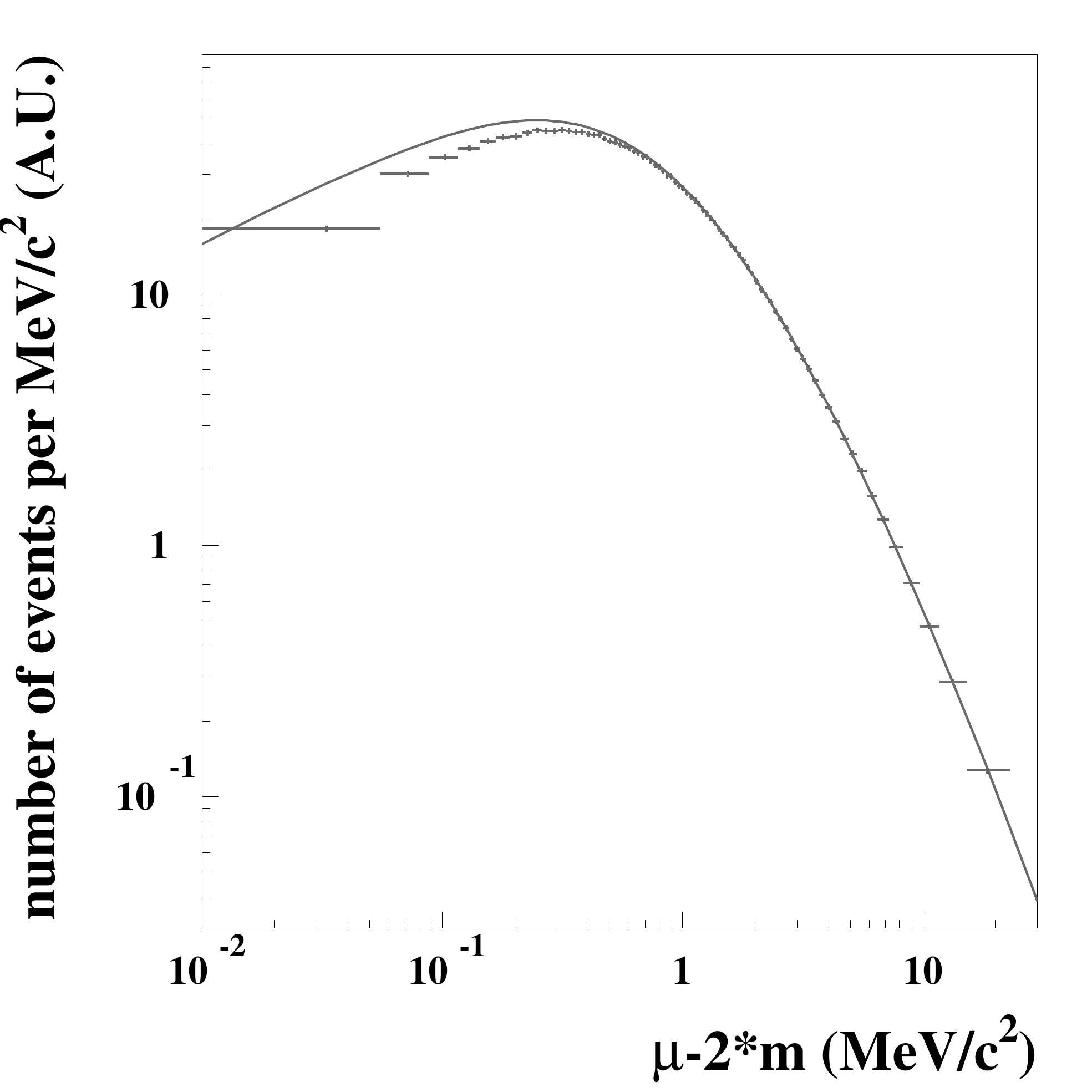}
\caption{Spectrum of the $e^+e^-$ invariant mass, minus its minimal 
 value of $2 m$, for nuclear conversions of 40\,GeV photons on silicon
(grey points: Monte Carlo sample;
 black curve: analytical expression from \cite{Borsellino:1953zz}).
The bin sizes have been set so that the numbers of events per bin are
identical.
The vertical ``error bar'' for each bin is located at the barycenter
of the events that enter that bin.
 \label{fig:mass}}
\end{center}
\end{figure}

Figure \ref{fig:mass} shows the comparison of the spectra obtained
with the 5D algorithm of \cite{Bernard:2018hwf} to
the low-mass approximation analytical expression \cite{Borsellino:1953zz},
for which 
the screening of the nuclear field by the atomic electrons
is described with a form factor
(eq. (10) of \cite{Borsellino:1953zz} or eq. (3.4) of \cite{Borsato:2015zuf}).
The simulated spectrum and the analytical expression are found to
agree with each other, except for the lowest invariant masses.
Borsato \cite{Borsato:2015zuf} had noticed a large discrepancy between
the fractions of events having an invariant mass
$\mu > 10\,\mega\electronvolt/c^2$ predicted by their Geant4
simulation (3.2\,\%) and computed from \cite{Borsellino:1953zz}
(5.4\,\%); we obtain a much better agreement with the 5D generator,
with a fraction found to be $(5.57 \pm 0.07) \% $.

Variations on the nature of the target (C, Si, Ge) or on the photon
beam energy (40\,GeV, 40\,TeV) induced very little variation of the
spectra, be it analytical or from simulated data (plots not shown).

\section{Polarization Asymmetry}

For gamma rays converting to an $e^+e^-$ pair, the measurement of the
fraction and of the angle of the linear polarization of is based on
the analysis of the distribution of an azimuthal angle, $\phi$,
describing the orientation of the final state
\cite{BerlinMadansky1950}

\begin{equation}
 \dd \sigma / \dd \phi \propto \left(1 + A \times P \cos(2\phi) \right)
 ,
 \label{eq:1D}
\end{equation}
\begin{itemize}
\item
$A$, the polarization asymmetry, depends on the energy, $E$, of the
photon and varies from $\pi/4$ at threshold 
\cite{Gros:2016dmp}
 (when the masses of the
particles of the pair are much smaller than that of the target,
compare Fig. 3 of \cite{Bernard:2019qvd}
to Fig. 5 of \cite{Bernard:2018hwf}) to
$1/7$ at very high-energy \cite{Boldyshev:1972va}.
Over most of the MeV-GeV energy range which is presently accessible to
experimentalists, $A$ is close to $0.15 - 0.20$ (see e.g. Fig. 3 of 
\cite{Gros:2016dmp}).
 
\item $P$ is the linear polarization fraction of the photon beam.
\end{itemize} 

Using the Weizsacker-Williams approximation, Wick has obtained
\cite{Wick1951} an expression for the polarized double-differential
cross section,
$\gfrac{\dd \sigma}{\sin{\theta_\ell} \dd \theta_\ell \dd \phi}$,
as a function of $\beta$, where $\beta$ and $\theta_\ell$ are, in the
pair center-of-mass frame, the velocity and the polar angle of either
of the leptons.

As the pair invariant mass, $\mu$, is related to $\beta$,
$\mu = 2 m / \sqrt{1 - \beta^2}$, we can obtain the
variation of the polarization asymmetry as a function of the pair
invariant mass $\mu$ \cite{Borsato:2015zuf}.
Wick's expression
integrates \cite{Wick1951} to
\begin{equation}
 \gfrac{ \dd \sigma }{\dd \phi} \propto \left(1 + \gfrac{1}{3} \cos^2{\phi}\right)
 ,
\end{equation}

that is 
\begin{equation}
 \gfrac{ \dd \sigma }{\dd \phi} \propto \left(1 + \gfrac{1}{7} \cos{2\phi} \right)
 ;
\end{equation}
that is $A = 1/7$, 
which is the high-energy limit of the Boldyshev-Peresunko high-energy
asymptotic expression obtained \cite{Boldyshev:1972va} from the
Bethe-Heitler differential cross section
\begin{equation}
A \approx \gfrac
{\gfrac{4}{9}\log{\left(2 E / m c^2\right)} - \gfrac{20}{28}}
{\gfrac{28}{9}\log{\left(2 E / m c^2\right)} - \gfrac{218}{27}}.
\label{eq:sig:HE}
\end{equation}

Integrating Wick's expression for
$\gfrac{\dd \sigma}{\sin{\theta_\ell} \dd \theta_\ell \dd \phi}$ on
$\theta_\ell$,
we obtain the polarization asymmetry as a function of $\beta$ 

\begin{equation}
 A = 
 \gfrac
 {\beta (\beta^2-1) + (1- \beta^4)\arctanh{\beta}}
 { \beta ( \beta^2 -2) + (3 - \beta^4) \arctanh{\beta}} 
 . \label{eq:Wick:int}
\end{equation}

\begin{figure}[ht!]
\begin{center}
\includegraphics[width=0.47\textwidth]{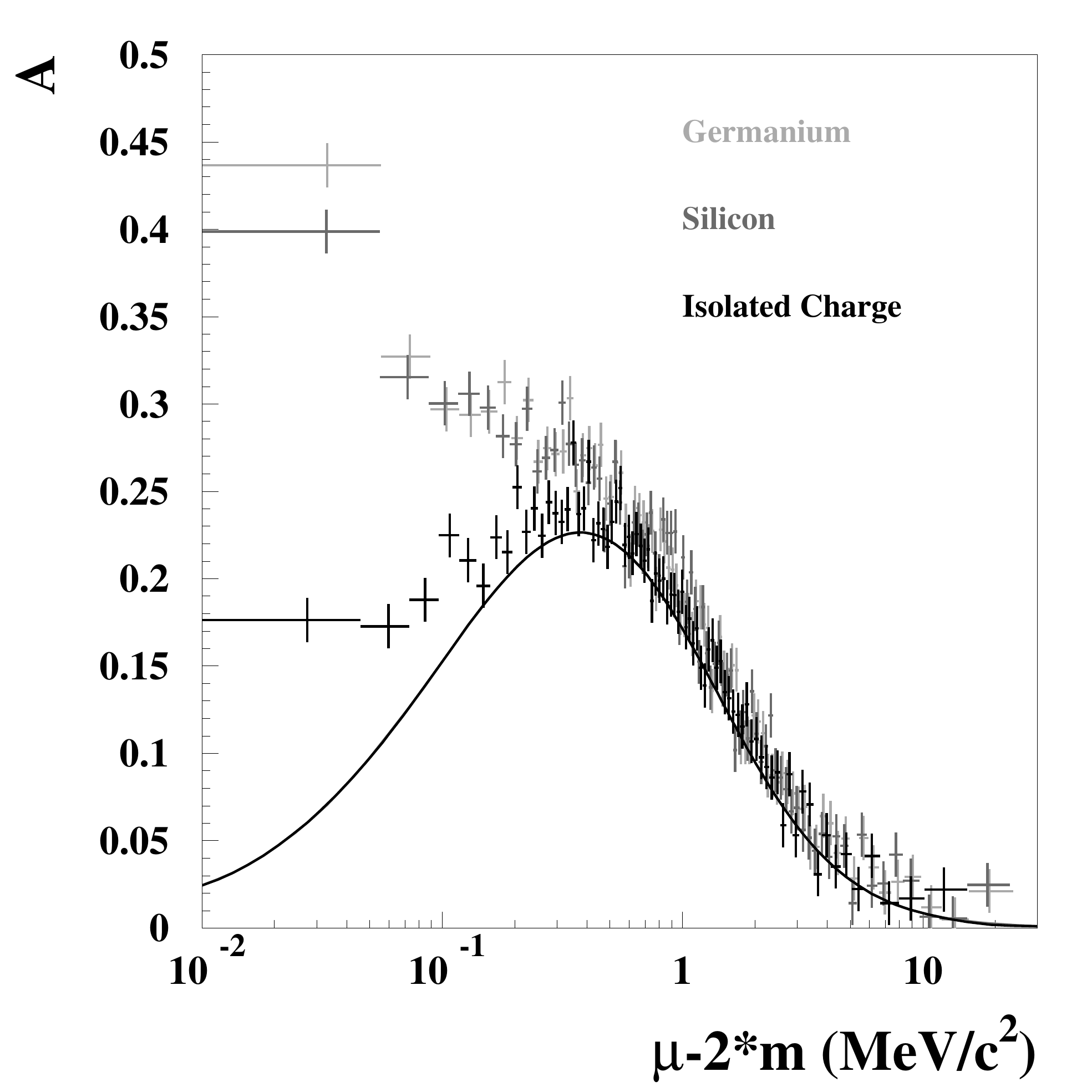}
\caption{Variation of the polarization asymmetry of the nuclear conversions of 40\,GeV photons,
 as a function of the dielectron invariant mass minus its minimal 
 value of $2 m$, for simulated samples
 (conversions on germanium (light grey), on silicon (grey) and
 on point-like charged targets (black)),
 compared to the analytical expression (eq. (\ref{eq:Wick:int})) obtained from Wick's differential cross section \cite{Wick1951}.
The bin sizes have been set so that the numbers of events per bin are
identical.
The vertical ``error bar'' for each bin is located at the barycenter
of the events that enter that bin.
 \label{fig:asy:mass}}
\end{center}
\end{figure}

We compute
the polarization asymmetry on simulated samples using
the moment's method \cite{Bernard:2013jea,Gros:2016dmp}, the azimuthal
angle of the event being defined as the bisectrix of the azimuthal
angles of the electron and of the positron \cite{Gros:2016dmp}.
For 40\,GeV photons converting on point-like charges we obtain
$A = (16.22 \pm 0.14)\%$, which is marginally compatible with the
Boldyshev-Peresunko \cite{Boldyshev:1972va} value of $A = 15.8\%$, while for conversions on
silicon we obtain $A = (16.70 \pm 0.14)\%$.
These values are significantly higher than Wick's high-energy limit of 
$A = 1/7 \approx 14.3\%$.

Figure \ref{fig:asy:mass} shows the variation of the polarization
asymmetry as a function of dielectron invariant mass for several targets.
On most of the mass range, say above $2 m + 0.5\,\mega\electronvolt/c^2$,
the shape of the polarization asymmetry for simulated data is well
described by eq. (\ref{eq:Wick:int}), with some underestimation
though, as we observed above on the whole sample.
At lower masses, simulated spectra are affected by the $q^2$
suppression due to screening in the case of conversions on full atom
(with their electron cloud),
where $q$ is the momentum of the recoiling target, 
and eq. (\ref{eq:Wick:int}) obtained from
Wick's model fails to represent the data.

\begin{figure}[ht!]
\begin{center}
\includegraphics[width=0.47\textwidth]{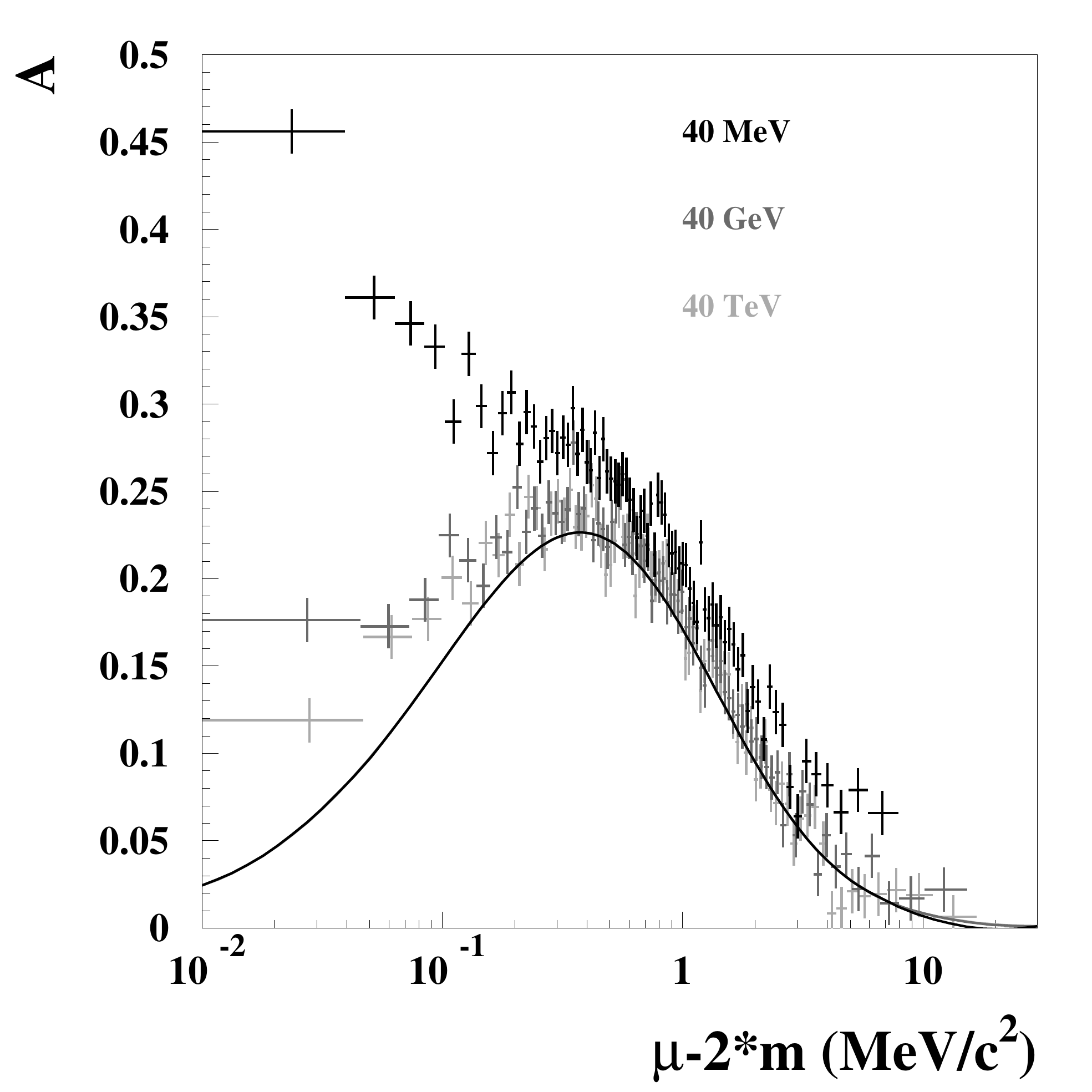}
\caption{Variation of the polarization asymmetry of nuclear
 conversions on silicon as a function of the dielectron invariant mass
 minus its minimal value of $2 m$, for simulated samples with various
 energies (40\,TeV (light grey), 40\,GeV (grey) and 40\,MeV (black)),
 compared to the analytical expression (eq. (\ref{eq:Wick:int}))
 obtained from Wick's differential cross section \cite{Wick1951}.
The bin sizes have been set so that the numbers of events per bin are
identical.
The vertical ``error bar'' for each bin is located at the barycenter
of the events that enter that bin.
 \label{fig:asy:mass:energy}}
\end{center}
\end{figure}

 Examination of the variation of the polarization asymmetry as a
 function of dielectron invariant mass for several incident photon
 energies (Fig. \ref{fig:asy:mass:energy}) confirms the interpretation of
 Wick's model as a high-energy approximation. 
 
\section{Conclusion}
\label{sec:Conclusion}

We have obtained the spectra of the dielectron invariant mass, $\mu$,
and the variation with $\mu$ of the polarization asymmetry, $A$, of
gamma-ray conversions to $e^+e^-$ pairs in the algorithm
\cite{Bernard:2018hwf} used in the five-dimensional
 generator of the Bethe-Heitler differential
cross section.
We have partly integrated Wick's expression of the differential cross
section to obtain the polarization asymmetry as a function of $\mu$
(eq. (\ref{eq:Wick:int})).
At large masses, the results on simulated data agree with Borsellino's
expression (spectrum) and with eq. (\ref{eq:Wick:int}) (polarization
asymmetry).
At smaller masses though, close to $\mu = 2 m$, Borsellino's and
Wick's expressions are found to fail to describe the simulated data.

Apart from the very small invariant mass range of
$\mu < 1\,\mega\electronvolt/c^2$ that is irrelevant for the analyses
performed in \cite{Aaij:2015dea,Borsato:2015zuf} and for which
form-factor effects affect the spectra and the polarization asymmetry,
simulated results are found to be in agreement with past published
analytical expressions.

\section{Acknowledgments}
\label{sec:Acknowledgments}

I'd like to express my gratitude to Marie-Hélène Schune who drew my
attention to the issue of the dielectron invariant mass spectrum in
gamma-ray conversions to pairs in past Geant4 physics models, in
relation with the analysis of polarization measurements in $b \to s
\gamma$ decays.

~

\end{document}